\documentclass{PoS}
\usepackage{epsfig}
\usepackage{multirow}

\newcommand{\tr}{\mbox{tr}}

\newcommand{\fm}{\mbox{fm}}
\newcommand{\sgn}{\mbox{sgn}}
\newcommand{\MeV}{\mbox{MeV}}
\newcommand{\GeV}{\mbox{GeV}}
\newcommand{\eq}{\begin{equation}}
\newcommand{\ee}{\end{equation}}
\newcommand{\ea}{\begin{eqnarray}}
\newcommand{\eea}{\end{eqnarray}}
\newcommand{\MSbar}{\overline{\mathrm{MS}}}
\newcommand{\RI}{\mathrm{RI'-MOM}}
\newcommand{\RGI}{\mathrm{RGI}}

\newcommand{\half}{\mbox{\small $\frac{1}{2}$}}      
\newcommand{\third}{\mbox{\small $\frac{1}{3}$}}      
\newcommand{\twelfth}{\mbox{\small $\frac{1}{12}$}}      

\newcommand{\Dlr}[1]{\stackrel{\leftrightarrow}{D^{#1}}}

\title{Nucleon Structure from Quenched Overlap Fermions}
\author{David Galletly$^a$, Martin G\"urtler$^b$, Roger Horsley$^a$, Karl Koller$^c$, Volkard Linke$^d$, Paul~E.L. Rakow$^e$, Charles J. Roberts$^e$,
Gerrit~Schierholz$^b$, \speaker{Thomas~Streuer}$^b$\\
\vspace*{0cm}\\
\llap{$^a$}University of Edinburgh \hfill DESY 05-177\\
Edinburgh EH9 3JZ, UK\hfill Edinburgh 2005/13\\
\llap{$^b$}John von Neumann-Institut f\"ur Computing NIC\\
Platanenalleee 6, 15738 Zeuthen, Germany\\
\llap{$^c$}Sektion Physik, Universit\"at M\"unchen\\
80333 M\"unchen, Germany\\
\llap{$^d$}Institut f\"ur Theoretische Physik, Freie 
Universit\"at Berlin\\
14196 Berlin, Germany\\
\llap{$^e$}Theoretical Physics Division, Department of 
Mathematical Sciences, University of Liverpool\\
Liverpool L69 3BX, UK\\
\\
E-mail:
\email{galletly@ph.ed.ac.uk},
\email{martin.guertler@desy.de},
\email{rhorsley@ph.ed.ac.uk},
\email{karl.koller@lrz.uni-muenchen.de},
\email{volkard.linke@physik.fu-berlin.de},
\email{rakow@amtp.liv.ac.uk},
\email{cjr@amtp.liv.ac.uk},
\email{gerrit.schierholz@desy.de},
\email{thomas.streuer@desy.de}\\
\\
{\bf QCDSF Collaboration}}

\abstract{
We compute the lowest moments of the nucleon's structure functions
using quenched overlap fermions at two different lattice spacings.
The renormalisation is done nonperturbatively in the $\RI$-scheme.
}

\FullConference{}
\FullConference{XXIIIrd International Symposium on Lattice Field Theory\\

                 25-30 July 2005\\

                 Trinity College, Dublin, Ireland}

\ShortTitle{Nucleon Structure}

\PoS{PoS(LAT2005)363}

\begin{document}

\section{INTRODUCTION}

Information about the internal structure of the nucleon is
encoded in its structure functions. While they cannot be computed directly on
the lattice, the operator product expansion (OPE) provides a connection between
their moments and nucleon matrix elements of local operators.
For instance, for the unpolarised structure function $F_1$,
the OPE reads
\eq
2\int_0^1{dxx^{n-1}F_1(x, Q^2) = \sum_f E^{(f)}_{F_1,n}v_n^{(f)}+O(1/Q^2)},
\ee
where $f$ denotes the quark flavour, $E^{(f)}_{F_1,n}$ is the (perturbative)
Wilson coefficient and the matrix element $v_n^{(f)}$ is defined by
\eq
\langle N(\vec{p})|O_{(f)}^{\{\mu_1 \ldots \mu_n \}}-\mathrm{traces}|
N(\vec{p})\rangle=2v_n^{(f)}(p^{\mu_1}\ldots p^{\mu_n}-\mathrm{traces}),
\ee
with the operator
\eq
O_{(f)}^{\mu_1\ldots\mu_n}=\bar{\psi}_{f}\gamma^{\mu_1}\Dlr{\mu_2}\ldots
\Dlr{\mu_n}\psi_{f}.
\label{eq_ops_unp}
\ee
Similar relations hold for the other structure functions, see e.g.
\cite{goeckeler1}
for details. Both the matrix element $v_n^{(f)}$ and the Wilson
coefficient $E^{(f)}_{F_1,n}$ depend upon the choice of a renormalisation
scheme and scale; only in their product, these dependencies cancel.

\section{LATTICE SIMULATION}

We use the overlap operator given by

\eq
D=\rho\big(1+\frac{m_q}{2\rho}+(1-\frac{m_q}{2\rho})
\gamma_5 \sgn(H_W(-\rho))  \big), 
\label{eq_D}
\ee
where $H_W(-\rho)=\gamma_5 (D_W-\rho)$, $D_W$ being the Wilson Dirac operator.
We approximate the
sign function appearing in (\ref{eq_D}) by minmax polynomials \cite{giusti1}.
For the gauge part we chose the L\"uscher-Weisz action \cite{luscher1}
\eq
S[U]=\frac{6}{g^2}\left( c_0
\sum_{\rm plaq}\third
\mbox{Re}\, 
\mbox{Tr}\, [1-U_{\rm plaq}] 
+c_1 \sum_{\rm rect}\third \mbox{Re}\, 
\mbox{Tr}\, [1-U_{\rm rect}]   
+c_2 \sum_{\rm par}\third 
\mbox{Re}\, \mbox{Tr}\, [1-U_{\rm par}] \right),
\label{eq_LW}
\ee
\vspace*{-0.4cm}

\begin{table}[b]
\centerline{
\begin{tabular}{|c|c|c|c|}
  \hline $V$ & $\beta$ & $a(\fm)$ & confs. \\\hline
  $16^332$ & $8.00$ & $0.153(3) \fm$ & 300\\
  $24^348$ & $8.45$ & $0.105(2) \fm$ & 200\\\hline
\end{tabular}
}
\caption{Parameters of the gauge configurations used.}
\label{table_configs}
\end{table}

\noindent
with coefficients $c_1$, $c_2$ ($c_0 + 8 c_1 + 8 c_2 = 1$) taken from tadpole 
improved perturbation theory~\cite{gattringer1}.
We ran our computations at two lattice spacings, see table \ref{table_configs}.
The scale has been set from the pion decay constant, details are given 
in \cite{martin}.
The parameter $\rho$ in (\ref{eq_D}) was 
set to $1.4$.

 In order to remove $O(a)$ errors from the
three-point functions, we employ the method of \cite{capitani1},
which amounts to replacing propagators $D^{-1}\Psi$ by
$\frac{1}{1-\frac{m}{2\rho}}D^{-1}\Psi-\frac{1}{2(1-\frac{m}{2 \rho})}\Psi$. 
Jacobi smeared point sources \cite{goeckeler1} with parameters \mbox{$N_s=50$} 
and \mbox{$\kappa_s=0.21$}
have been used in order to obtain a good overlap with the ground state.

The computation of matrix elements follows the procedure outlined in
\cite{goeckeler1}:
We form the ratio
\eq
R=\frac{\langle N(t_{sink})O(\tau)\bar{N}(t_{source}) \rangle}
{\langle N(t_{sink})\bar{N}(t_{source})\rangle},
\label{eq_ratio}
\ee
from which the matrix element can be extracted in the region
$t_{source} < \tau < t_{sink}$. We always set $t_{source}=0$ and
$t_{sink}=9$ ($t_{sink}=13$) on the $\beta=8.0$ ($\beta=8.45$) configurations
(in lattice units), which corresponds to a
distance between source and sink of $1.4~\fm$.

The matrix elements we are considering are listed in table \ref{table_ops},
along with the operators used for their determination,
where we use the operators (\ref{eq_ops_unp}) and
\eq
O_{5,(f)}^{\mu_1\ldots\mu_n}=\bar{\psi}_{f}\gamma_5\gamma^{\mu_1}\Dlr{\mu_2}\ldots
\Dlr{\mu_n}\psi_{f}.
\ee
We compute
only flavour non-singlet matrix elements, since in this case there is no
contribution from disconnected diagrams.

\begin{table}[b]
\centerline{
\begin{tabular}{|c|c|}
  \hline Matrix Element & Operator \\\hline
  $g_A$ & $\bar{\psi}\gamma^5\gamma^2\psi$ \\
  $g_T$ & $\bar{\psi}\gamma^5\sigma^{24}\psi$ \\
  $v_2$ & $O^{44}-\third\left(O^{11}+O^{22}+O^{33} \right)$ \\
  $a_1$ & $\half\left(O_5^{24}+O_5^{42}\right)$ \\\hline
\end{tabular}
}
\label{table_ops}
\caption{Operators used in the three-point functions.}
\end{table}

\section{NON-PERTURBATIVE RENORMALISATION}

The operators appearing inside the three-point functions have to be
renormalised. For $g_A=\Delta u-\Delta d$, the operator to be used is the axial current $A_\mu$,
the renormalisation of which is particularly simple because it does not
depend on a renormalisation scheme or scale. 
It can be obtained from a Ward identity~\cite{giusti2} as
\eq
Z_A=\lim_{m_q \rightarrow 0} \lim_{t \rightarrow \infty}\frac{2m_q}{m_\pi}
\frac{\langle P(t)P(0)\rangle}{\langle A_4(t)P(0)\rangle}.
\label{eq_ZA}
\ee

The renormalisation constants of the other operators under consideration 
are logarithmically divergent.
We have computed them in the $\RI$-scheme~\cite{martinelli1}.
 In this scheme, the
renormalisation condition is formulated in terms of quark Greens functions
in Landau gauge with an operator insertion at zero
momentum transfer:
\eq
C_O(p)=\frac{1}{V}\sum_{x,y,z}{e^{-ip(x-y)}
\langle \psi(x)O(z)\bar{\psi}(y) \rangle}.
\ee
From this quantity, the amputated vertex function $\Gamma_O$ is formed:
\eq
\Gamma_O(p)=S^{-1}(p)C_O(p)S^{-1}(p),
\ee
with the quark propagator
\eq
S(p)=\frac{1}{V}\sum_{x,y}{e^{-ip(x-y)}\langle \psi(x)\bar{\psi}(y)\rangle }.
\ee
The renormalisation condition at scale $\mu$ is
\eq
\left. Z_\psi(\mu) Z_O(\mu)\Pi_O\left(\Gamma_O(p)\right)\right|_{p^2=\mu^2}=1,
\ee
with the projector $\Pi_O(\Gamma)=\twelfth\tr\left(\Gamma_{O,\mathrm{Born}}^{-1}(p)\Gamma \right)$.
The wavefunction renormalisation constant $Z_\psi$ has been
determined from the relation $Z_\psi Z_A \Pi_A\left(\Gamma_A\right)=1$.

In order to convert the results to the $\MSbar$-scheme, we first determine
the renormalisation group invariant renormalisation constant $Z_O^{\RGI}$:
\eq
Z_O^{\RGI}=\left(Z_O^{\RI,\RGI}(\mu)\right)^{-1}Z_O^{\RI}(\mu),
\label{eq_rgi}
\ee
and then convert to the $\MSbar$-scheme at scale $\mu'$
(we always use $\mu'=2\GeV$):\\
$Z_O^{\MSbar}(\mu')=Z_O^{\MSbar,\RGI}(\mu)Z_O^{\RGI}$
with the conversion functions
\eq
Z^{{\cal S},\mathrm{RGI}}_O(\mu)=
\left(2b_1{g^{\cal S}(\mu)}^2\right)^{-\frac{d_{O,1}}{2b_1}}
		   \exp{\left[\int_0^{g^{\cal S}(\mu)}d\xi
		     \left(
		     \frac{\gamma^S_O(\xi)}{\beta^S(\xi)}
		     +\frac{d_{O,1}}{b_1\xi}
		     \right)
		   \right]}.
\ee
The coefficients of the $\beta$ and $\gamma$ functions are taken from \cite{gracey1, gracey2}.

\begin{figure}[t]
\begin{center}
\epsfig{file=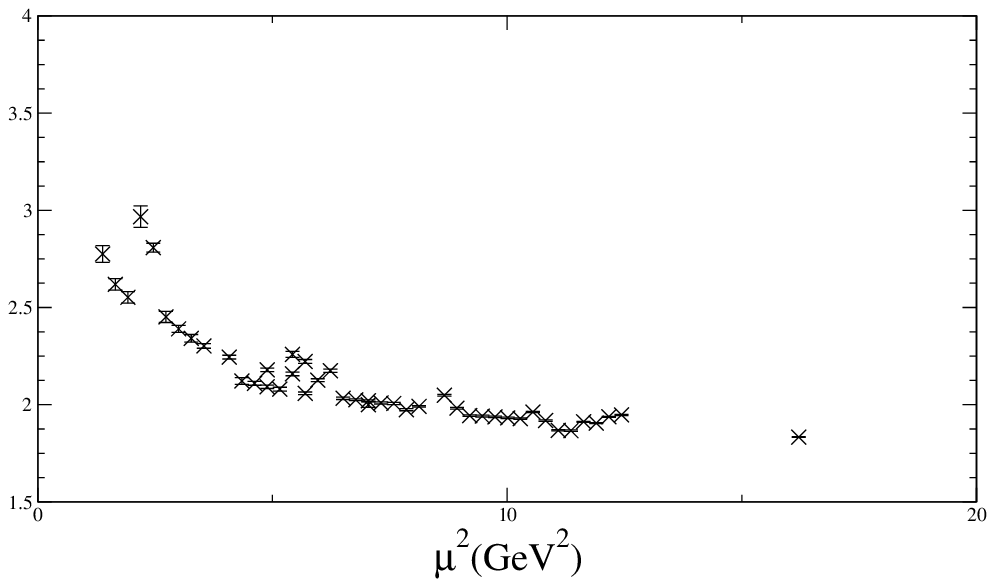, width=7cm, height=4.5cm}
\epsfig{file=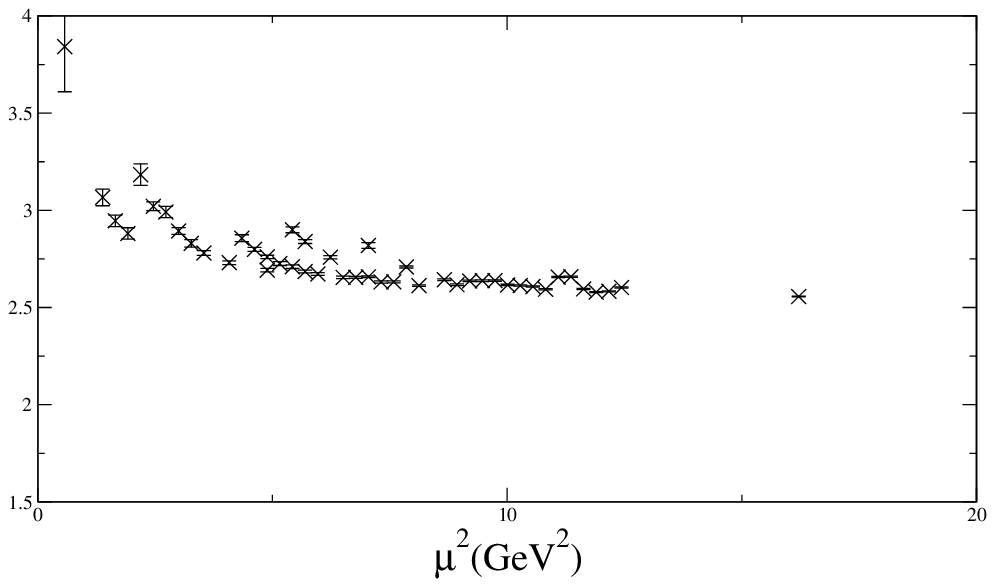, width=7cm, height=4.5cm}
\caption[]{Left: The renormalisation constant $Z_{v_{2b}}^{\RI}(\mu)$,
Right: The same, but with the renormalisation group running removed
according to (\ref{eq_rgi}).}
\label{plot_Zv2b}
\end{center}
\end{figure}

In fig. \ref{plot_Zv2b} we plot $Z_{v_{2b}}^{\RI}$ and
$Z_{v_{2b}}^{\RGI}$ for $\beta=8.45$. From the latter plot, we read off 
$Z_{v_{2b}}^{\RGI}=2.6$ from the plateau region $8\GeV^2<p^2<15\GeV^2$.
Using $Z_{v_{2b}}^{\MSbar,\RGI}(2\GeV)=0.737$, we obtain
\mbox{$Z_{v_{2b}}^{\MSbar}=1.92$}. The renormalisation constants for all operators we
need are shown in table \ref{table_Z}. A comparison with results obtained
in one-loop tadpole-improved lattice perturbation theory \cite{horsley1, horsley2} shows large
discrepancies, especially for the operators with one derivative.

\begin{table}[b]
\centerline{
\begin{tabular}{|c|c|c|c|c|}
  \hline \multirow{2}{*}{Operator} & \multicolumn{2}{c|}{$\beta=8.0$}
& \multicolumn{2}{c|}{$\beta=8.45$} \\\cline{2-5}
 &pert. & nonp. & pert. & nonp. \\\hline
  $O_{A_\mu}$  & 1.36  &  $1.59$  & $1.30$  & $1.42$\\
  $O_{g_T} $   & 1.36  &  $1.73$  & $1.33$  & $1.54$\\
  $O_{v_2} $   & 1.33  &  $2.11$  & $1.39$  & $1.92$\\
  $O_{a_1} $   & 1.34  &  $2.21$  & $1.40$  & $1.98$    \\\hline
\end{tabular}
}
\label{table_Z}
\caption{Comparison between the renormalisation constants $Z_O^{\MSbar}(2\GeV)$
obtained non-perturbatively and in lattice perturbation theory.}
\end{table}

\section{RESULTS}

\begin{figure}[t]
\begin{center}
\epsfig{file=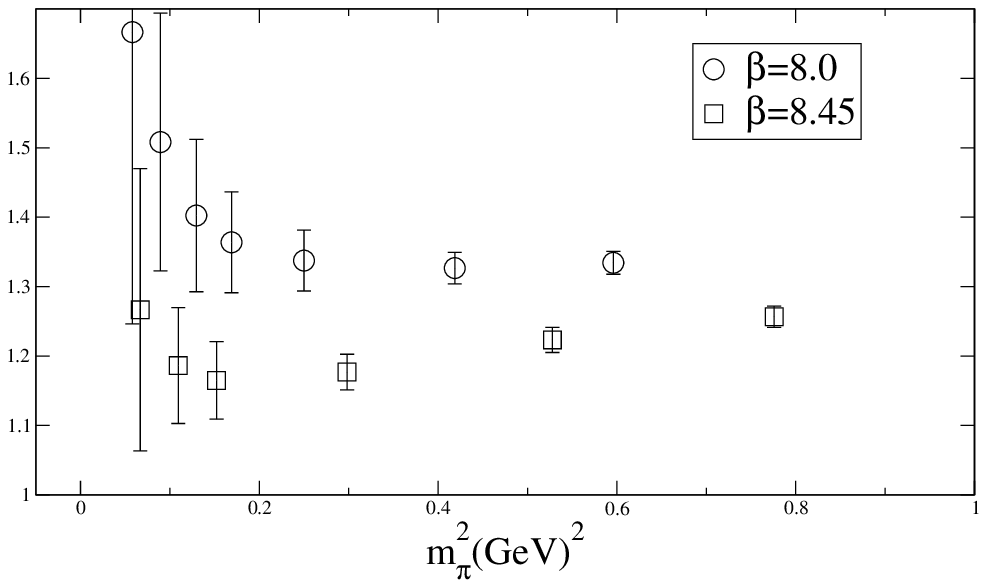, width=7cm, height=4.5cm}
\epsfig{file=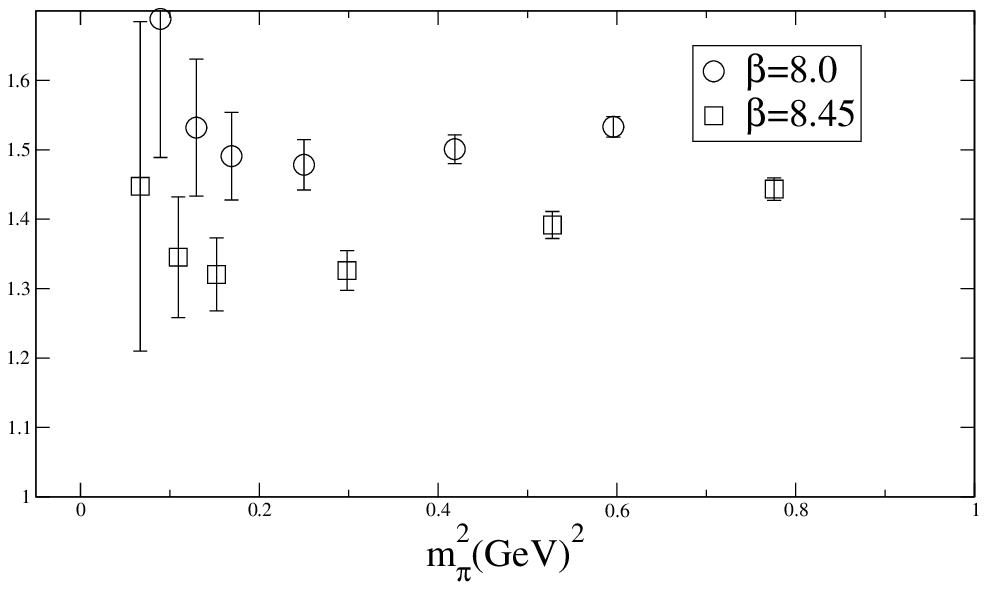, width=7cm, height=4.5cm}
\caption{The nucleon's axial charge (left plot) and tensor charge (right plot)
as a function of the squared pion mass}
\label{plot_gah1}
\end{center}
\end{figure}

Our results for the axial charge $g_A$ are displayed in fig. \ref{plot_gah1} (left).
 A linear extrapolation to the
chiral limit yields $g_A=1.37(5)$ at $\beta=8.0$ and 
$g_A=1.14(5)$ at $\beta=8.45$.
The tensor charge $g_T=\delta u-\delta d$ is plotted in
fig. \ref{plot_gah1} (right); in this case, 
linear extrapolations to $m_q=0$ yield $g_T=1.35(4)$ at $\beta=8.0$ and
$g_T=1.18(5)$ at $\beta=8.45$.

\begin{figure}[t]
\begin{center}
\epsfig{file=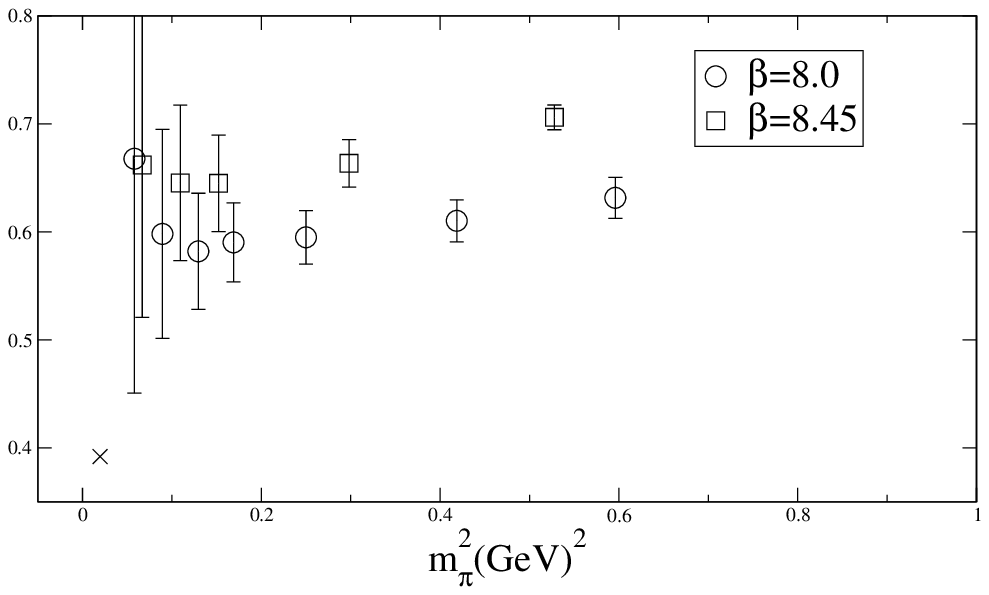, width=7cm, height=4.5cm}
\epsfig{file=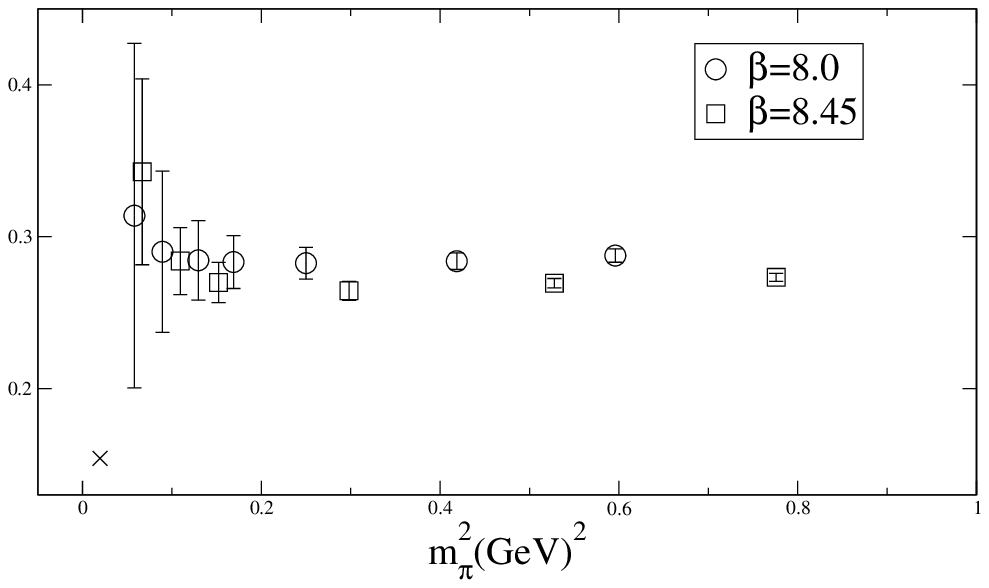, width=7cm, height=4.5cm}
\caption{The nucleon matrix elements $a_1^{u-d}$ (left plot) and $v_2^{u-d}$
(right plot) as functions of the squared pion mass ($\MSbar$, $2\GeV$).}
\label{plot_a1v2}
\end{center}
\end{figure}

The results for the matrix elements $v_2^{u-d}=\langle x \rangle^{u-d}$ and
 $a_1^{u-d}=2\langle x \rangle^{\Delta u-\Delta d}$ are shown in fig.
\ref{plot_a1v2}, together with the phenomenological values.
In both cases, there is almost no dependence on the quark
mass visible. In the range $m_\pi \gtrsim 400\MeV$, our results
agree with previous results obtained from improved Wilson fermions~\cite{goeckeler1}.

\section{CONCLUSIONS}

We have determined the flavour non-singlet nucleon matrix
elements $g_A$, $g_T$, $v_2$ and $a_1$ from quenched overlap fermions.
The renormalisation has been done nonperturbatively. Comparing
the results at the two values for the lattice spacing we have,
we find significant discretisation effects, in contrast with the situation
for hadron masses~\cite{martin}.\\

While our results are in good agreement with earlier determinations,
there remains a rather large discrepancy to the phenomenological values
for $v_2$ and $a_1$, even at the lowest quark masses we can reach at present.


\section*{ACKNOWLEDGEMENTS}

The numerical calculations were performed at the HLRN
(IBM pSeries 690), at NIC J\"ulich (IBM pSeries 690) and at the PC farms
at DESY Zeuthen and LRZ Munich. 
We thank these institutions for their support.
Part of this work is supported by DFG under contract FOR 465 (Forschergruppe Gitter-Hadronen-Ph\"anomenologie)



\end{document}